# The Journal Coverage of Web of Science, Scopus and Dimensions: A Comparative Analysis

Vivek Kumar Singh[1#], Prashasti Singh[1], Mousumi Karmakar[1],
Jacqueline Leta[2] & Philipp Mayr[3]

[1]Department of Computer Science, Banaras Hindu University, Varanasi-221005, India.
[2]Institute of Medical Biochemistry, Federal University of Rio de Janeiro, Brazil.
[3] GESIS-Leibniz Institute for the Social Sciences, Cologne, Germany.

**Abstract:** Traditionally, Web of Science and Scopus have been the two most widely used databases for bibliometric analyses. However, during the last few years some new scholarly databases, such as Dimensions, have come up. Several previous studies have compared different databases, either through a direct comparison of article coverage or by comparing the citations across the databases. This article aims to present a comparative analysis of the journal coverage of the three databases (Web of Science, Scopus and Dimensions), with the objective to describe, understand and visualize the differences in them. The most recent master journal lists of the three databases is used for analysis. The results indicate that the databases have significantly different journal coverage, with the Web of Science being most selective and Dimensions being the most exhaustive. About 99.11% and 96.61% of the journals indexed in Web of Science are also indexed in Scopus and Dimensions, respectively. Scopus has 96.42% of its indexed journals also covered by Dimensions. Dimensions database has the most exhaustive journal coverage, with 82.22% more journals than Web of Science and 48.17% more journals than Scopus. This article also analysed the research outputs for 20 selected countries for the 2010-2018 period, as indexed in the three databases, and identified database-induced variations in research output volume, rank, global share and subject area composition for different countries. It is found that there are clearly visible variations in the research output from different countries in the three databases, along with differential coverage of different subject areas by the three databases. The analytical study provides an informative and practically useful picture of the journal coverage of Web of Science, Scopus and Dimensions databases.

**Introduction**

Traditionally, research assessment exercises in scientometric studies have more frequently drawn research output data from one of the two well-known scholarly databases, namely- Web of Science or Scopus. The Web of Science and Scopus databases have been extensively compared, both through direct comparisons of coverage (such as Gavel & Iselid, 2008; Mongeon & Paul-Hus, 2016) as well as through citation-comparison studies (such as Martin-Martin et al., 2018). These and several other previous studies on comparison of Web of Science and Scopus have largely indicated that Scopus has a wider coverage of journals as compared to Web of Science, whereas Web of Science continues to be more selective in its journal coverage. The Google Scholar database, which came up in the same year (2004) as that of Scopus, has also been used for different bibliometric studies, however, less commonly for research assessment exercises. (see for example Mayr & Walter, 2007; Falagas et al., 2008; De Winter, Zadpoor & Dodou, 2014; Harzing & Alakangas, 2016). During the last few years, several new databases, such as Microsoft Academic Search and Dimensions, have attracted wide attention of the bibliometrics research community. With emergence of these new databases, many studies (such as Visser, van Eck & Waltman, 2019,2020; Harzing, 2019; Martin-Martin et al., 2020) attempted to compare the article coverage of these bibliographic data sources with the existing databases.

---
[#] Corresponding Author. Email: vivek@bhu.ac.in

Among the newer scholarly databases, we find the Dimensions database as one of the most serious contenders of the Web of Science and Scopus databases. A recent study (Thelwall, 2018) has also indicated that Dimensions database could be an alternative to existing databases. However, since Dimensions database includes preprints as well, which means that citations from not yet peer reviewed articles are also included, which creates the theoretical opportunity for manipulative citation gaming. Hence, Thelwall (2018) has suggested its use for traditional research evaluation with the above rider. We explored different features of the Dimensions database and found two quite interesting features which may eventually make it as a competitor to Web of Science and Scopus, for research assessment and evaluation exercises as well. *One*, it provides a set of "filters" that can limit database search results to a set of journals, such as nationally recognized journal lists. This implies that Dimensions may be able to have a readymade mechanism through which evaluators from a specific country can evaluate the research performance of the different institutions in their country as per their own country journal lists. *Second*, it has an article-level subject classification system, unlike the source-based classification of Web of Science and Scopus. This may make classification of articles into different subject areas relatively more accurate and hence subject area specific research performance evaluations more informed. Further, Dimensions has a deliberate design feature of multiple classifications which allows it to integrate '*n*' classifications, including alignment with different national science classification systems. It is in this context that we have attempted to compare the journal coverage of Dimensions along with a fresh comparison of the journal coverage of the two already established databases- Web of Science and Scopus. Though the article-level comparative analysis approach, adopted by many earlier studies can be more comprehensive, the journal-level coverage analysis has its own relevance, for example, for studies on Bradford Law or studies on citation and research evaluation, each of which are directly influenced by the number of journals indexed in a database.

Some recent studies compared different aspects of several old and new scholarly databases, including Dimensions. Thelwall (2018) compared Dimensions with Scopus by analysing a random sample of 10,000 Scopus indexed articles in Food Science. Harzing (2019) compared several databases, including the Dimensions database by using two data samples- one of a single academic and another of selected journals. Visser et al. (2019) compared article coverage of Web of Science, Scopus, Dimensions and Crossref through a publication record match in the entire collection of articles in the databases. Martin-Martin et al. (2020) compared six data sources including Dimensions through a citation-based indirect comparison approach. Visser, van Eck & Waltman (2020) performed a direct pair-wise comparison of article coverage of Scopus database with several other databases, including Dimensions. However, to the best of our knowledge, no previous study has compared the journal coverage of Dimensions with Web of Science and Scopus. Further, the most recent existing study on journal coverage comparison of Web of Science and Scopus by Mongeon & Paul-Hus (2016) has used master journal lists of 2014 and hence has become outdated. This study (Mongeon & Paul-Hus, 2016) has also shown variations in research output and rank of the 15 highly productive countries. However, given that it used data only up to the year 2014, it would be interesting to revisit this aspect as well with the updated data. Thus, this article attempts to present an alternative/ complementary approach of analysing the journal coverage of the three databases (Web of Science, Scopus and Dimensions). We use direct comparison of overlapping and unique journals covered in the three databases (both pair-wise and all three databases taken together) and also analysed what impact the coverage variations may have on research output volume, rank and global share of different countries. Differences in subject area composition of the data collected from the three databases is also explored.

**A brief overview of the three scholarly databases**

This section presents a very brief overview of some important information about the three scholarly databases- Web of Science, Scopus and Dimensions. Among these databases, Web of Science is the oldest database, created in the year 1964, followed by Scopus in the year 2004, and Dimensions being the most recent database, created in the year 2018. For a more detailed overview of different bibliographic data sources, one may refer to the special issue of Quantitative Science Studies[2].

*Web of Science*

Web of Science is the oldest among the three scholarly databases, created initially as an information retrieval tool in 1964 by Eugene Garfield from Institute of Scientific Information (ISI). It was called Science Citation Index (SCI), initially covered 700 journals, and was primarily meant to be a citation index. Over a period of time, it has grown and added new citation indices, the Social Sciences Citation Index (SSCI) in 1973, the Arts & Humanities Citation Index (AHCI) in 1978, and a Book Citation Index (BKCI) in 2011. The SCI, SSCI and AHCI were combined together and launched on the World Wide Web as Web of Science in 1997. Another citation index called Emerging Sources Citation Index (ESCI) was launched in 2015 with an objective to provide early visibility for titles being evaluated for inclusion in their classical indices- SCIE, SSCI, and AHCI (Somoza-Fernández, Rodríguez-Gairín, & Urbano, 2018).

Currently, Web of Science is owned by Clarivate Analytics[3]. As per the latest data[4] of 2020, the Web of Science Core Collection covers more than 74.8 million scholarly data and datasets, 1.5 billion cited references (dating back to 1900) across 254 subject-disciplines. The Science Citation Index Expanded (SCIE) indexes 9,200 journals across 178 scientific disciplines comprising of total 53 million records and 1.18 billion cited references; the SSCI indexes 3,400 journals across 58 social sciences disciplines comprising of total 9.37 million records and 122 million cited references; and the AHCI indexes 1,800 journals across 28 Arts & Humanities disciplines comprising of total 4.9 million records and 33.4 million cited references. Despite a huge growth in publication sources over the years, SCIE, SSCI & AHCI have been selective in coverage, with indexing decisions made at regular intervals. If a journal gets indexed by one of the citation indices, all its articles get covered in Web of Science. The Web of Science web interface provides different kinds of data search and download facilities. The downloaded data for basic usage has about 68 field tags, such as PT (for Publication Type), OI (for Orcid

---

[2] https://www.mitpressjournals.org/toc/qss/1/1
[3] https://clarivate.com/webofsciencegroup/
[4] https://clarivate.com/webofsciencegroup/solutions/web-of-science-core-collection/

Identifier), DT (for Document Type) etc. A recent article by Birkle et al. (2019) provides a detailed overview of Web of Science database.

*Scopus*

Scopus database, created in 2004, is a product of Elsevier. It is often considered as one of the largest curated databases covering scientific journals, books, conference proceedings etc., which are selected through a process of content selection followed by continuous re-evaluation. The decision to index a journal or conference or other publication is taken by a Content Selection and Advisory Board (CSAB). When it was launched in 2004, it contained about 27 million publication records for the period 1966 to 2004. Currently, it covers publication records from the year 1788 onwards, with approximately 3 million records added every year. The recently updated (Oct. 2019) content coverage guide[5] of Scopus shows that, it comprises of about 23,452 active journal titles, 120,000 conferences and 206,000 books from more than 5,000 international publishers. However, the master journal list of Scopus has entries for higher number of journals, some of which are no longer active. Unlike, Web of Science, it has a single citation index, covering journal and conference articles in different subject areas. The Scopus content coverage guide indicates that it contains a total of about 77.8 million core records.

Scopus platform allows data access by Search, Discover and Analyse options. The Search option allows document, author and advanced search. The Discover option enables users to identify collaborators, research organizations with respect to research output, finding related publication data through various metrics such as author keywords, shared references etc. The Analyse option is a tool to track citations, assessment of search results on criteria such as country wise, affiliation wise, research area wise distribution of resultant data. The data downloaded from Scopus database for research publications usually comprises of 43 fields, such as abbreviated source title, abstract, author keywords, source title (journal in which it is published), document type etc. A more detailed overview of Scopus database could be found in Baas et al. (2019).

*Dimensions*

Dimensions is the newest among the three scholarly databases, created in 2018. Dimensions provides a single platform access to different kinds of research data, as also indicated in its official description, which says "*it brings together data describing and linking awarded grants, clinical trials, patents and policy documents, as well as altmetric information, alongside traditional publication and citation data*" (Herzog, Hook & Konkiel, 2020). In the launch version of 2018, it contained about 90 million publications (of which 50 million had full-text versions) with more than 873 million citation links (Hook, Porter & Herzog, 2018). As in Sep. 2019, it contained 109 million publications indexed with about 1.1 billion citations. In addition to publications and citations, it also contains other linking information about clinical trials (497,000), patents (39 million), policy documents (434,000), and the altmetric data points (111 million). In terms of journals covered, as on 1st April. 2019, Dimensions had publication records from more than 50,000 journals (Bode et al., 2019). However, a recently shared journal list (May 2020) by Dimensions contains more than 74,000 journal entries, which shows that Dimensions is a new and growing database covering many smaller publishers as well.

Unlike Web of Science and Scopus, Dimensions uses a different approach for sourcing data, with Crossref and PubMed being the "*data spine*". The *bottom-up* approach used indicates that the data sourced from Crossref and PubMed is further enhanced by collecting data about affiliations, citations etc. The data enhancements are done through a data enhancement process that takes data from various sources such as DOAJ, initiatives like Open Citations and I4OC, clinical trial registries, openly available public policy data, and other Digital Science companies like Altmetric and IFI Claims. Dimensions database, being the most recent and drawing data from different sources, has quite rich data about institution ids, grant ids etc. Dimensions can be accessed in three forms- Dimensions, Dimensions Plus and Dimensions Analytics, with different levels of privileges attached to each. The data downloaded from web interface of Dimensions for research publications usually comprises of 71 different fields, such as title, source title (journal in which the research paper is published), authors affiliations, country of research organization etc. A more detailed overview of Dimensions database could be seen in Herzog, Hook & Konkiel (2020) and some recent analytical studies on Dimensions database could be found in (Thelwall, 2018; Martin-Martin et al., 2018; Harzing, 2019; Visser, van Eck & Waltman, 2019).

---

[5] https://www.elsevier.com/__data/assets/pdf_file/0007/69451/Scopus_ContentCoverage_Guide_WEB.pdf

**Related Work**

There are several previous studies that compared different scholarly databases. These studies have either taken a direct approach of comparing the journals/articles indexed in different databases or compared differences in citations across different databases for a given sample of articles. We present here a brief report of some of the most relevant previous studies.

*Studies on direct coverage comparison of databases*

Some of the initial studies of direct coverage comparison of different databases are Mayr & Walter (2007), Gavel & Iselid (2008), Lopez-Illescas et al. (2008), Lopez-Illescas et al. (2009) and Vieira & Gomes (2009). Mayr & Walter (2007) compared a German journal list in Social Sciences to find out which journals were also indexed in Google Scholar. Gavel & Iselid (2008) analysed the overlaps between Web of Science (WoS) and Scopus and some other major scientific databases. They observed that Scopus surpasses WoS in STM areas (Science, Technology and Management), but has a limited humanities coverage for the pre-1996 era. They further observed that overlaps of journal titles in various citation databases differ in coverage and consistency. Lopez-Illescas et al. (2008) performed a comparison of the journal coverage of Web of Science and Scopus in the field of Oncology and observed that Web of Science was subset of Scopus, with Scopus covering 90% more oncological journals than Web of Science. They also discovered a strong correlation between the classical Web of Science impact and SCImago Journal Rank (SJR). Lopez-Illescas et al. (2009) performed a comparative analysis among journals included in Web of Science and Scopus in the field of Oncology, focusing on country rankings in terms of the published article count and average citation impact. They found that the additional oncological journals in Scopus mainly served the national research fraternity and that inclusion of additional journals in Scopus benefitted countries in terms of published article count but showed a declining trend in terms of average citation rate. Vieira & Gomes (2009) worked on coverage and overlap between Web of Science and Scopus for university domain (for a set of Portuguese universities) for the year 2006. They concluded that about 2/3$^{rd}$ of the documents referenced in any of the two databases may be found in both databases while a fringe of 1/3$^{rd}$ is only referenced in one or the other. They also observed that citation impact of the documents in the core present in both databases is usually higher.

Chadegani et al. (2013) attempted to perform a comprehensive comparison of Scopus and Web of Science databases, mainly to answer questions about similarity and differences in title coverage of the two databases. It was found that Web of Science had a strong coverage which goes back to 1990 and most of its journals covered were written in English. However, Scopus covered a superior number of journals but with lower impact and limited to recent articles. Mongeon & Paul-Hus (2016) analysed the journal coverage of Web of Science and Scopus, mainly to assess whether some fields, publishing countries and languages were over or underrepresented. The coverage of active scholarly journals in WoS (13,605 journals) and Scopus (20,346 journals) was compared with Ulrich's extensive periodical directory (63,013 journals). Analytical results indicated that the use of either WoS or Scopus for research evaluation may introduce biases that favour Natural Sciences and Engineering as well as Biomedical Research to the detriment of Social Sciences and Arts & Humanities. Similarly, English-language journals were found overrepresented to the detriment of other languages.

Harzing & Alakangas (2016) compared Web of Science, Scopus and Google Scholar longitudinally for eight data points between 2013 and 2015 on a sample of 146 academicians in five broad disciplinary areas. They observed a consistent and reasonably stable quarterly growth for both publications and citations across the three databases. AlRyalat, Malkawi & Momani (2019) analysed coverage, focus and tools in popular databases- PubMed, Scopus and Web of Science by using data for publications from Jordanian authors in the years 2013-2017.They observed that PubMed focuses mainly on life sciences and biomedical disciplines, whereas Scopus and Web of Science are multidisciplinary. Aksnes & Siversten (2019) analysed relative coverage of publication type, field of research and language of Scopus and Web of Science by using data for Norwegian scientific and scholarly publication output in 2015 and 2016. The results showed that Scopus covered 72% of the total publications, whereas the Web of Science Core Collection covered 69%. The coverages are found higher in medicine and health (89% in Scopus & 87% in WoS) and in natural sciences and technology (85% in Scopus & 84% in WoS) as compared to social sciences (48% in Scopus & 40% in WoS) and humanities (27% in Scopus & 23% in WoS).

*Studies on comparing citation counts in databases*

Yang & Meho (2006) compared citations found in Scopus and Google Scholar with those found in Web of Science, for items published by two Library and Information Science full-time faculty members. They observed that Web of Science should not be used alone for locating citations to an author or title and that Scopus and Google Scholar can help identify a considerable number of valuable citations not found in Web of Science. Google Scholar was found to have several technical problems in accurately and effectively locating citations. Falagas et al. (2008) compared the citations in PubMed, Google Scholar, Scopus and Web of Science in the field of biomedical research. They observed that Scopus offers about 20% more coverage than Web of Science, whereas Google Scholar offers results of inconsistent accuracy. Bar-Illan (2008) compared the h-indices of a list of highly-cited Israeli researchers based on citations counts retrieved from the Web of Science, Scopus and Google Scholar. They observed that the results obtained through Google Scholar are considerably different from the results based on the Web of Science and Scopus. They found large variations in the h indices of researchers in different databases and observed that the h-index and citation counts do not follow a linear relationship.

Torres-Salinas, Lopez-Cózar & Jiménez-Contreras (2009) compared citation differences in Web of Science and Scopus, by analysing data in the area of Health Sciences of the University of Navarra in Spain. They found that papers received 14.7% more citations in Scopus than in Web of Science. The difference in citations did not correspond to the difference of coverage of the two databases. Mingers & Lipitakis (2010) analysed the citations that a paper receives, as per Web of Science and Google Scholar. They compared dataset of over 4,600 publications from three UK Business Schools. The results show that Web of Science is indeed poor in the area of management and that Google Scholar, while somewhat unreliable, has a much better coverage. They concluded that Web of Science should not be used for measuring research impact in management. Adriaanse & Rensleigh (2011) performed a macro- and micro-level comparison of the citation resources- Web of Science (WoS), Scopus and Google Scholar (GS)- for the environmental sciences' scholarly journals in South Africa during 2004-2008. The macro-level evaluation results indicated that Scopus surpassed both WoS and GS whereas the micro-level evaluation results indicated that WoS surpassed both Scopus and GS. In another study, Adriaanse & Rensleigh (2013) compared three citation resources- ISI Web of Science, Scopus and Google Scholar- with one another to identify the citation resource with the most representative South African scholarly environmental sciences citation coverage. It was found that WoS performed the best as it retrieved the most unique items, followed by Google Scholar and Scopus.

De Winter & Dimitra Dodou (2014) analysed the development of citation counts in Web of Science (WoS) and Google Scholar (GS) for two classic articles and 56 articles from diverse research fields, making a distinction between retroactive growth and actual growth. Results showed that GS had substantially grown in a retroactive manner, whereas retroactive growth of WoS was small. The actual growth percentages were moderately higher for GS than for WoS. Martin-Martin et al. (2018) investigated 2,448,055 citations to 2,299 English-language highly-cited documents from 252 Google Scholar (GS) subject categories published in 2006, comparing GS, the Web of Science (WoS) Core Collection, and Scopus. The results suggested that in all areas GS citation data is essentially a superset of WoS and Scopus, with substantial extra coverage. Google Scholar found nearly all citations found by WoS (95%) and Scopus (92%), and a large number of unique citations. It was observed that about half of Google Scholar unique citations were not from journals. In another study, Martín-Martín, Orduna-Malea & López-Cózar (2018) performed an analysis of 2,515 highly-cited documents published in 2006 that Google Scholar displayed in its classic papers and checked whether they were present in Web of Science and Scopus and with similar or varying citation counts. The results showed that a large fraction of highly-cited documents in the Social Sciences and Humanities were invisible to Web of Science and Scopus. In the Natural, Life, and Health Sciences the proportion of missing highly-cited documents in Web of Science and Scopus were much lower.

*Studies on comparison of newer scholarly databases*

Thelwall (2018) is the first among the recent studies that explored Dimensions database from an impact assessment perspective, choosing a random sample of 10,000 Scopus articles from 2012 in Food Science research during 2008–2018. The results indicated high correlations between citation counts from Scopus and Dimensions (0.96 by narrow field in 2012) as well as similar average counts. Almost all Scopus articles with DOIs were found in Dimensions (97% in 2012). Thus, the study concluded that the scholarly database component of Dimensions seemed to be a plausible alternative to Scopus and the Web of Science for general citation analyses and for citation data.

Harzing (2019) explored the new (free) sources for academic publication and citation data along with well-established Google Scholar, Scopus and the Web of Science databases. Microsoft Academic (2016), Crossref (2017) and Dimensions (2018) were the newer databases studied. The study tried to investigate the full publication and citation record of a single academic, as well as six top journals in Business & Economics. The small-scale study suggests that, when compared to Scopus and the Web of Science; Crossref and Dimensions have a similar or better coverage for both publications and citations, but a significantly lower coverage than Google Scholar and Microsoft Academic. The study found Google Scholar and Microsoft Academic as the most comprehensive free sources for publication and citation data.

Huang et al. (2020) have explored differences across data sources and their implications for metrics and rankings at institutional scale. They performed bibliographical comparison between Web of Science, Scopus and Microsoft Academic at the institution level for two sets of 15 and 140 universities, however most of the detailed analysis was only for 15 universities. They used a DOI based comparison of the three data sources and concluded that "*the coverage of DOIs not only differ across the three sources, but their relative coverages are also non-symmetrical*". They inferred that use of just one database could seriously disadvantage some institutions in assessments. They further observed that Microsoft Academic has greater coverage than Scopus and Web of Science.

Visser et al. (2019) compared coverage of Web of Science, Scopus, Dimensions and Crossref through a publication record match in the entire collection of articles in the databases. They found that Dimensions had higher coverage as compared to Web of Science and Scopus. They observed that Dimensions covered about 78% of the publication records available in Scopus. In a later study, Visser, van Eck & Waltman (2020) performed a large-scale comparison of five multidisciplinary bibliographic data sources: Web of Science, Scopus, Dimensions, Crossref, and Microsoft Academic. They used a direct comparison approach and considered all scientific documents from the period 2008-2017 covered by these data sources, except few document types excluded from the comparison. They have compared Scopus in pair-wise manner with all the other data sources, and analysed differences in coverage of different data sources, with focus on differences over time, differences per document type, and differences per discipline. They obtained following key conclusions: (a) Scopus covers a large number of documents not covered by WoS whereas almost all journal articles covered by WoS are also covered by Scopus, (b) Scopus covers a large number of journal articles that are not covered by Dimensions and Crossref, and the other way round, Dimensions and Crossref cover an even larger number of journal articles that are not covered by Scopus, (c) Microsoft Academic has perhaps the most comprehensive coverage of scientific literature. Finally, they proposed that "*there is value both in the comprehensiveness offered by Dimensions and Microsoft Academic and in selectivity offered by Scopus and Web of Science*" and that "*comprehensiveness and selectivity are no longer mutually exclusive*".

Martin-Martin et al. (2020) analysed 3,073,351 citations found by the six data sources (Google Scholar, Microsoft Academic, Scopus, Dimensions, Web of Science, and OpenCitations' COCI) to 2,515 English-language highly cited documents published in 2006 from 252 subject categories. They found that Google Scholar is the most comprehensive source, with almost 88% of all citations covered by it. They, however, found that in many subjects Microsoft Academic and Dimensions are good alternatives to Scopus and Web of Science in terms of coverage. For example, Microsoft Academic found 60% of all citations, including 82% of Scopus citations and 86% of Web of Science citations. Dimensions had a little less than Scopus citations covered, with 84% of Scopus citations and 88% of Web of Science citations. They finally concluded that for comprehensive citation counts (without complete list of citing sources), Google Scholar is the best choice. However, if a complete list of citing sources is needed then Microsoft Academic is the best choice.

**Data & Methodology**

Two kinds of data have been used in the analysis: (a) the master journal lists of the three databases, and (b) the publication records in the databases, for the period 2010-2018, for twenty selected countries.

The master journal lists of Web of Science and Scopus are publicly available on their websites and were updated in June 2020. Dimensions database does not have a publicly available master journal list, indicating the list of journals for which it may be covering content. Therefore, we requested the Dimensions team to provide us with a master journal list. The Dimensions team provided a curated list of the journals covered, as updated till May 2020. The master journal list of Web of Science was provided in three separate files, one each for journals indexed in SCIE, SSCI and AHCI. The three lists were updated till June 2020. We have downloaded all these three files and later merged them together to create a single master journal list comprising of journals indexed in the Web of Science core collection (SCIE, SSCI and AHCI taken together). The combined master journal list of Web of Science had 13,610 entries, each of which comprised of following fields: journal title, ISSN, e-ISSN, Publisher name, Publisher address, Languages, and Web of Science Categories. The master journal list of Scopus had 40,385 entries, each of which included the following fields: Source record id, Source Title, Print-ISSN, e-ISSN, Active or Inactive, Coverage period, article language expressed as three ISO language codes, CiteScore, Publisher's Name, Publisher's Country/Territory, All Science Journal Classification Codes (ASJC) etc. The Dimensions comprehensive journal list contained 77,471 entries, each of which comprised of the following fields: DOI, id, title, ISSN, e-ISSN, publisher name. Table 1 presents the number of journals covered in each of the three databases, along with total publication record counts in the three databases for document types 'article' and 'review' for the period 2010-18.

In addition to the master journal lists, we have also obtained publication records for twenty selected countries from all the three databases. Since different databases had different number of publication records for various countries, we selected the twenty countries having highest number of publication records in Web of Science database. The ordering of Web of Science was preferred as it is the oldest and still one of the most-well-known multidisciplinary sources of scientific information. The purpose of obtaining data for these countries was to observe variations in publication records for the same countries in the three databases. Analysis of publication records from the countries with high outputs was suitable for a clear illustration of the variations in publication record counts of the different countries across the three databases. Another major issue was that different databases categorise the publication records in different document types. For example, table 2 shows the total counts for publication records of different document types in the three databases for the period 2010-18. As we can see, the publication records are categorized into various document types. Therefore, for a direct and meaningful comparison, we have used only publication records for the document types 'article' and 'review' only. It may be noted that Dimensions does not have document type 'review' and apparently review articles published in journals are indexed as document type 'article'. The publication record counts for 'article' and 'review' document types for the twenty selected countries were obtained through advanced search in web-interfaces of Web of Science and Scopus and through API queries from Dimensions. For Web of Science, search query PY=

(2010-2018) was used with data limited to twenty selected countries to get the publication counts. In case of Scopus, the corresponding publication record counts were obtained through a query of the form PUBYEAR>2009 AND PUBYEAR<2019. For Dimensions database, the corresponding data was obtained by using API queries with publication year filters set to the range 2010-2018.

The data for master journal lists as well as publication records for twenty selected countries was then computationally analysed. The analysis comprised of three main parts. *First*, the master journal lists of the three databases were compared to identify overlapping and uniquely indexed journals. The comparison of master journal lists of the three databases comprised of a set of pre-processing and matching steps. The pre-processing steps involved removing duplicate and incomplete entries. The matching steps involved comparing the journal entries in the three lists to identify overlapping and unique journals. The detailed description of pre-processing and matching steps is provided in the **Appendix**.

*Secondly*, the research output rank, global share and compounded annual growth rate (CAGR) for the twenty selected countries were computed. The objective here was to assess the impact of varied coverage of journals by the three databases on the research output of different countries as indexed in the three databases. The publication record counts for the twenty selected countries for the period 2010-18 were obtained and analysed. The research output rank, global share and compounded annual growth rate (CAGR) for the twenty selected countries were computed. For a direct comparison, only the data for document types 'article' and 'review' was used for the purpose.

*Thirdly*, the subject area distribution of publication records for the three databases was analysed. It may be noted that the three databases have publication records classified under different major and minor subject areas. Therefore, it was not possible to directly compare the subject area distribution of publication records in the databases. We tried to obtain a correspondence between different subject area classes by mapping the subject area of Scopus and Dimensions databases to the best matching major subject area of the Web of Science database. Table 3 describes the subject area class mapping of Scopus and Dimensions to Web of Science major subject areas. The Web of Science database has Life Sciences, Physical Sciences, Technology, Social Sciences, and Arts & Humanities as major subject areas. The twenty-seven macro subject areas of Scopus were mapped to these five major subject areas of Web of Science. Similarly, the twenty-two subject areas of Dimensions were mapped to the closest matching Web of Science subject area. This mapping allowed us to do a broad-based comparison of the subject area distribution of research output across the three databases. One may point out here that the Dimensions database uses an article-level subject classification, unlike the source-based subject classification used by Web of Science and Scopus. Therefore, mapping the subject areas of Dimensions to Web of Science major subject areas is equivalent to mapping an article-level subject area classification to a source-based classification. While, this may not be the best strategy, but there was no other manageable alternative available as one would need a very detailed and comprehensive computational effort to classify publication records in Web of Science and Scopus on article-level. Further, a previous study (Singh, Piryani, Singh, & Pinto, 2020) have shown that the article-level subject area classification of Dimensions may not necessarily be practically more accurate than source-based classification of Web of Science and Scopus. Therefore, we decided in favour of these mappings to Web of Science major subject areas so that at least a broad-based comparison of the subject area distribution of the publication records can be done.

**Results**

The analytical results are organized in three parts. *First of all*, the overlapping and unique journals in the three databases are identified. *Secondly*, the research output for twenty selected countries obtained from the three databases is analysed to understand database-induced variations in research output volume, relative rank and global share of the twenty selected countries. *Thirdly*, the subject area distribution of research output data of the twenty selected countries from the three databases are analysed to understand the variation in the subject area distribution of the three databases.

*Journal coverage of the three databases*

The master journal lists of Web of Science (comprising of 13,610 journal entries), Scopus (comprising of 40,385 journal entries), and Dimensions (comprising of 77,471 journal entries) were compared to identify overlaps in the journal coverage of the three databases. Figure 1 shows a Venn diagram representation of these pair-wise overlaps of journals indexed in the three databases. It is seen that 13,489 journals are common in Web of Science and Scopus; 13,149 journals are common in Web of Science and Dimensions; and 38,336 journals are common in Scopus and Dimensions.

The pair-wise journal overlaps of Web of Science with the two other databases could be summarized as follows:

Web of Science- total pre-processed journal entries= 13,610
- Overlap with Scopus- 13,489 (99.11% of Web of Science), 121 journals are non-overlapping (0.89% of Web of Science)
- Overlap with Dimensions- 13,149 (96.61% of Web of Science), 461 journals are non-overlapping (3.39% of Web of Science)
- Unique journals in Web of Science- 19 (0.14% of Web of Science)

The pair-wise journal overlaps of Scopus with the two other databases could be summarized as follows:

Scopus- total pre-processed journal entries= 39,758
- Overlap with Web of Science- 13,489 (33.93% of Scopus), 26,269 journals are non-overlapping (66.07% of Scopus)
- Overlap with Dimensions- 38,336 (96.42% of Scopus), 1,422 are non-overlapping (3.58% of Scopus)
- Unique journals in Scopus- 980 (2.46% of Scopus)

The pair-wise journal overlaps of Dimensions with the two other databases could be summarized as follows:

Dimensions- total pre-processed journal entries= 73,966
- Overlap with Web of Science- 13,149 (17.78% of Dimensions), 60,980 are non-overlapping (82.22% of Dimensions)
- Overlap with Scopus- 38,336 (51.82% of Dimensions), 35,630 are non-overlapping (48.17% of Dimensions)
- Unique journals in Dimensions- 35,528 (48.03% of Dimensions)

In addition, to pair-wise overlaps, we have also computed the journal overlap in all the three databases taken together. Figure 2 shows a Venn diagram representation of overlaps and unique journals in the three databases. A total of 13,047 journals are found to be covered by all the three databases. This constitutes 95.86% of Web of Science, 32.82% of Scopus and 17.64% of Dimensions.

*Analysing research output of different countries in the three databases*

Since the three databases showed a significant degree of variation in their journal coverage, therefore, we tried to analyse what impact it may have on research output volumes, rank and global share of different countries. The research output corresponding to document types 'article' and 'review' for 20 selected countries was analysed. Given that Dimensions has been found to be most exhaustive in coverage of journals, we expected that the research output volumes of all the twenty countries may be highest in Dimensions database. Similarly, the research output volume was expected to be lesser in Web of Science as compared to both Scopus and Dimensions.

Table 4 shows research output rank, volume, and global share of the twenty selected countries for the period 2010-18. The research output volume for the same country is found to vary across the three databases. These variations are as large as 1 million records in several cases, such as for USA (between Web of Science and Dimensions) and China (between Web of Science and Scopus). In addition to research output volume, it was observed that the global share for different countries also varied across the databases, more so between Web of Science & Dimensions and Scopus & Dimensions. For example, USA is seen to have global share of 27.24% in research output as per Web of Science databases, whereas Scopus data shows its global share as 23.04% and Dimensions data shows the global share to be 17.03%. Thus, the variation in global share as per the three databases is as large as 10% in case of USA alone. China is also seen to have variations in global share, with 17.26% in Web of Science, 17.80% in Scopus and 9.38% in Dimensions. In case of India, the global share is found to be 4.00% in Web of Science, 4.72% in Scopus, and 2.74% in Dimensions. The research output ranks of different countries were also found to vary across the three databases. For example, India has research output rank of 9$^{th}$ in Web of Science, 6$^{th}$ in Scopus, and 7$^{th}$ in Dimensions. Thus, India has higher rank in Dimensions, although its global share in Dimensions is lesser than that of Web of Science. The relative ranks of several other countries were found to be different in the three databases. For example, countries like Canada and Italy, Spain and Australia, have different relative ranks as per Web of Science and Scopus databases. Many other countries are ranked in different orders in the three databases. The Web of Science and Scopus, however, show higher agreement in rank and global share of different countries as compared to Web of Science & Dimensions and Scopus & Dimensions.

Figure 3 shows a bar plot of the research output volume (in millions) of the twenty selected countries in the three databases for the 2010-18 period. It is observed that for most of the countries, the research output volume in Dimensions is higher than Scopus, which in turn is higher than in Web of Science. China and India are two major exceptions, where research output volume in Scopus is significantly higher than in Dimensions. This is an interesting observation, particularly given the fact that Dimensions covers almost twice the number of journals as compared to Scopus. There could be two reasons for this: (a) Dimensions may be having a preferential coverage of journals from the developed countries, and/ or (b) China and India may be producing less research in certain subject areas which may have significant (or higher) journal coverage in Dimensions. We look into the varied subject area coverage aspect in the next section.

Table 5 shows the year-wise research output of the twenty selected countries along with the CAGR value for research output in each of the three databases. The CAGR values for the same countries are found a bit different in data from the three databases. For example, USA has CAGR of 2.63% in Web of Science but 4.03% in Dimensions; China has CAGR of 8.85% in Scopus but 13.17% in Web of Science; India has CAGR of 6.51% in Web of Science but 10.59% in Dimensions. In most of the cases, Dimensions shows a higher CAGR as compared to Web of Science and Scopus for the same set of countries. These variations in research output volume, rank, global share and CAGR values thus provide an indication of existence of some demographic variation too in the coverage of the three databases. These variations may also be due to the fact that different countries may have varying research strengths in different subject areas.

*Subject area distribution of research output in the three databases*

In order to better understand the variations in research output volume, rank and global share of different countries across the three databases, as observed in the previous section, we tried to analyse the subject area distribution of publication records in the three databases.

First, the whole research output data for all the twenty countries taken together is analysed in terms of its subject area distribution. The major subject areas of Web of Science database are taken as reference subject areas, as described in the Data & Methodology section. Figure 4 shows the subject area distribution of the combined research output of the twenty selected countries in all the three databases. The major subject area distribution of Web of Science is found to comprise of 44.5% publication records in Life Sciences, 22.8% in Physical Sciences, 24.4% in Technology, 6.7% in Social Sciences and 1.3% in Arts & Humanities. In case of Scopus, the major subject area distribution includes 45.2% publication records in Life Sciences, 22.6% in Physical Sciences, 20.6% in Technology, 8.7% in Social Sciences and 2.7% in Arts & Humanities. In Dimensions database, the major subject area distribution includes 43.3% publication records in Life Sciences, 19.2% in Physical Sciences, 19.4% in Technology, 14.4% in Social Sciences, and 3.5% in Arts & Humanities. It is observed that Web of Science and Scopus in general have higher share of publications in Physical Sciences and Technology areas. Further, all the three databases have Life Sciences as the most prominent disciplinary area, with a share of more than 40% output. Dimensions has higher proportion of output in Social Sciences and Arts & Humanities than both Web of Science and Scopus. We can see that Dimensions has about 14.4% of its output in Social Sciences as compared to 8.7% in Scopus and 6.7% in Web of Science. Similarly, Dimensions has 3.5% of its output in Arts & Humanities, whereas Scopus has 2.7% of its output and Web of Science has 1.3% of its output in Arts & Humanities. Therefore, it appears that Dimensions may be having a better coverage of research output in Social Sciences and Arts & Humanities major subject areas. This could possibly also explain why China and India have lesser research output volume in Dimensions as compared to other countries, since both these countries have significantly lesser research output in Social Sciences and Arts & Humanities.

Secondly, we looked at subject area distribution of research outputs of all the twenty countries individually. Figure 5 shows this subject area distribution for all the twenty selected countries in the three databases. It is observed that most of the countries have 40-50% share of publications in the area of Life Sciences. China, Russia and Taiwan are, however, major exceptions. Further, Countries like Germany, Japan, France and India are also found to have higher proportion of research output in the Physical Sciences area. Similarly, China, India, South Korea and Taiwan are all found to have good proportion of research output in the Technology subject area. China and India have negligible research output in Arts & Humanities across all the databases and very low research output in Social Sciences. It is also observed that research output proportions of different subject areas for the same country also vary across the three databases. Thus, the results indicate that different databases not only vary in terms of their overall journal coverage but also in terms of coverage of journals from different subject areas.

**Discussion**

This article attempts to compare the three databases in terms of their journal coverage, both by comparing them pair-wise and by taking all the three databases together. The analytical results are new and useful in following ways:

- *First*, it provides most updated direct comparison of journal coverage of Web of Science and Scopus, with the most recently updated master journal lists (updated till June 2020). The last known direct comparison of journal coverage of the two databases was done by Mongeon & Paul-Hus (2016) which used the master journal lists of 2014. The databases have expanded in their journal coverage since then. Therefore, this study provides an up-to-date account of the journal coverage overlaps of the two databases.

- *Second*, the study is the first attempt towards direct comparison of journal coverage of Dimensions database with Web of Science and Scopus, which in turn helps in understanding how exhaustive or selective Dimensions is in its journal coverage as compared to Web of Science and Scopus. Further, the subject area distribution of the coverage of the three databases is also analysed and results indicate that Dimensions has relatively better coverage in the areas of Social Sciences and Arts & Humanities.

The previous study by Mongeon & Paul-Hus (2016) analysed the coverage overlap of Web of Science and Scopus by discipline, country and language of journals as per the information provided in Ulrich's periodical database. It was observed that for majority of the disciplines, Scopus includes most of the journals indexed in Web of Science. The journal overlaps varied across disciplines, with about 35% to 50% of the journals indexed in Scopus being also covered in Web of Science. It was also found that Scopus has a larger number of exclusive journals than Web of Science in all fields. Our analytical results show that about 34% of journals indexed in Scopus overlap with Web of Science and that Scopus has about 66% of its journals exclusively covered as compared to Web of Science. Therefore, the difference in journal coverage of Web of Science and Scopus is found to have grown with time.

The recent study by Visser, van Eck & Waltman (2020) compared article coverage of several bibliographic data sources, including Web of Science, Scopus and Dimensions. They have found that for the period 2008-17, Web of Science has about 22.9 million articles indexed, Scopus has about 27 million articles indexed and Dimensions has about 36.1 million articles indexed. The study compared the overlap across different databases with Scopus as baseline. It was observed that Scopus had about 17.7 million articles (about 65% of its 27 million articles) overlapping with Web of Science and about 21.3 million articles (about 78% of its 27 million articles) overlapping with Dimensions. Our analysis shows that in terms of journal coverage, Scopus has about 34% of its journals overlapping with Web of Science and about 96% of its journals overlapping with Dimensions. Thus, the journal coverage overlap of Scopus with Web of Science and Dimensions is different from article coverage percentage levels observed in the study by Visser, van Eck & Waltman (2020). It would be relevant here to mention that their study used data of articles for the period 2008-17, whereas Dimensions, being a new database, has grown substantially after 2017/2018. Moreover, different journals publish different number of articles, therefore, journal overlap levels need not necessarily be similar to article overlap levels of the databases. The overall conclusion of this previous study seems to remain valid in our results as well. We observe that Web of Science is the most selective, Scopus covers much larger number of journals than Web of Science, and Dimensions is significantly larger in coverage as compared to both the databases.

The recent study by Martin-Martin et al. (2020) compared six data sources through comparison of citations and observed that as far as citations are concerned, Microsoft Academic Search and Dimensions are a good alternative to Scopus and Web of Science in terms of coverage. They analysed 3,073,351 citations to 2,515 highly-cited articles and found that Dimensions covered 84% of Scopus citations and 88% of Web of Science citations. Dimensions found more citations than Scopus in 36 subject categories and more citations than Web of Science in 185 subject categories. They suggested that in use cases where exhaustiveness of coverage is required, in presence of large coverage divergence, a combination of several bibliographic data sources may be used. Our study has also shown that none of the databases completely cover the journals indexed in any other database. Each database has at least some uniquely covered set of journals, though it is small for Web of Science and larger for Scopus and Dimensions.

Our study also observed that differences in journal coverage of the three databases result in variation in research output volumes, rank and global share of different countries. Given that Dimensions has bigger coverage than Scopus and Web of Science, we expected that the research output volumes of the countries may follow the same pattern (smallest in Web of Science and largest in Dimensions) and that their relative ranks and global share may be similar across the databases. However, we observed that the twenty selected countries not only differ in research output volume across the three databases, but their relative research output ranks and global shares also vary across the three databases. Interestingly, some countries show higher research output in Scopus as compared to Dimensions. Moreover, some variations are found in the relative ranks of different countries in different databases. This clearly indicates that the databases have varied coverage of journals across geographies. Mongeon & Paul-Hus (2016) also observed these variations in rank and research output volumes of 15 highly productive countries in the data from Web of Science and Scopus.

The variation in differential coverage of several countries (for example China and India have higher research output in Scopus as compared to Dimensions) prompted us to explore the angle of differences in coverage of different subject areas by different databases. China and India have their main research strengths in Science and Technology areas, with relatively lesser research output in Social Sciences and Arts & Humanities. Interestingly, they are found to have lesser research output in Dimensions than Scopus, though most of the other countries have higher research output in Dimensions as compared to Scopus. This may be an indication that Dimensions has a better coverage of Social Sciences and Humanities journals. Our analytical results of comparing the subject area distribution of articles of the twenty selected countries confirmed this observation. The subject area distribution of combined research output of the twenty selected countries shows a higher proportion of Social Sciences and Arts & Humanities in Dimensions as compared to Web of Science and Scopus. Thus, it can be said that Web of Science and Scopus have more or less similar coverage across different subject areas, but Dimensions database provides a better coverage of journals in the areas of Social Sciences and Arts & Humanities.

**Conclusion**

This study compared the journal coverage of the three databases- Web of Science, Scopus and Dimensions, identifying what number of journals are commonly and uniquely covered by the three databases. The study complements the previous studies on direct comparison of databases at the article-level, by performing a journal coverage analysis, and has provided an informative and practically useful account of the journal coverage of Web of Science, Scopus and Dimensions databases, and its impact. The analytical study obtains following main conclusions.

- *First*, the three databases are found to differ significantly in their journal coverage, with Web of Science having the most selective journal coverage, whereas Dimensions having the most exhaustive journal coverage. It is found that almost all journals indexed in Web of Science are also covered by Scopus and Dimensions. Scopus indexes 66.07% more unique journals as compared to Web of Science and Dimensions covers 82.22% and 48.17% more unique journals as compared to Web of Science and Scopus, respectively.

- *Second*, the varied journal coverage of the three databases results in variation in research output volume, rank and global share of different countries. Therefore, drawing data from different databases may produce different outcomes for any bibliometric evaluation exercises done at the level of countries.

- *Third*, the three databases also vary in their coverage of different subject areas. The Web of Science and Scopus have majority of their coverage in Life Sciences, Physical Sciences and Technology Area. On the other hand, Dimensions appears to have a significantly better coverage of Social Sciences and Arts & Humanities.

The study shows that the three databases are at different extremes on the scale of exhaustivity and selectivity. Web of Science continues to be selective, whereas Dimensions provides a much wider and exhaustive coverage. One may like to use a particular database depending on the purpose of use. The Dimensions database, in particular, appears to be a promising source due to its exhaustive coverage and provision of data filters (which may limit the data to expert-curated or nationally recognized journal lists).

**Limitations**

This study has few limitations. *First*, the study used a somewhat restrictive and conservative matching procedure, which tried to avoid false positives. Therefore, there may be few cases of the same journals spelled differently in the different lists which could not be captured through our matching procedure. However, the impact of such cases on the overall results is likely to be very small, as observed in our manual scans of some random samples. *Second*, the study only focused on the analysis of journal coverage of different databases and analysis of other publication types such as - books and conferences - were out of the scope of comparison. It may be an interesting future work to analyse how much the three databases agree or differ in coverage of books and conferences. Further, the study has not included Emerging Sources Citation Index (ESCI) of Web of Science, as it is not part of Web of Science core collection and impact factors of journals in it are not published. This may be taken up as a future work.

# APPENDIX

This appendix describes the detailed steps of pre-processing and matching steps applied to the master journal lists from the three databases.

**Pre-processing:**

On a detailed inspection, the three master journal lists were found to have some duplicate and incomplete entries. Further, since the Dimensions database also includes preprints and conferences, its comprehensive journal list contained some entries that referred to preprint servers or conference proceedings. Therefore, the journal lists were pre-processed to remove duplicate and incomplete entries and entries for preprint servers and conference proceedings. The pre-processing steps applied are as follows:

Pre-processing step 1: In the first pre-processing step, we analysed journal entries on two keys: ISSN and e-ISSN. In each of the journal lists, entries that had both these fields null were removed first. Thereafter, entries that had both ISSN and e-ISSN fields duplicated were removed. Thus, at the end of pre-processing step 1, we were left with 13,610 entries in Web of Science journal list (out of total 14,737 entries), 39,851 entries in Scopus journal list (out of total 40,385 entries), and 74,705 entries in Dimensions journal list (out of total 77,471 entries).

Pre-processing step 2: In the second step, we analysed inconsistent entries where same ISSN or e-ISSN values occurred in different journal entries. Such entries (with repeated ISSN or e-ISSN values) were identified and removed. The Web of Science journal list had no such entry. In Scopus, 93 such duplicate occurrences were found and removed, leaving the remaining list to comprise of 39,758 entries. In Dimensions, 112 such entries were found and removed, and the remaining list comprised of 74,593 entries.

Pre-processing step 3: In the third step, the entries in the journal lists have been checked to see if they contain any entry for a non-journal publication source. It was found that the Dimensions journal list included some entries for preprints and conference proceedings as well. Accordingly, the journal list entries were scanned to see occurrence of certain keywords, such as preprint, preprints, preprint-server, symposium, conference, congress etc. A total of 7 entries were found for preprint sources in Dimensions list and were removed. A total of 617 entries were found for conferences in Dimensions list and were removed. The resulting journal list of Dimensions database contained 73,966 journal entries.

Thus, the pre-processed journal list of Web of Science contained 13,610 journal entries; Scopus pre-processed list had 39,758 journal entries, and Dimensions pre-processed list contained 73,966 journal entries.

**Matching:**

After the pre-processing steps, a systematic matching process was used to identify overlapping and unique journal records in different databases. We used a step-by-step matching process which used simple matching in the initial steps and a more restrictive matching strategy in later steps when the remaining journal lists became smaller. In the beginning, we did an ISSN/ e-ISSN based record matching and then later on used title-based and title text similarity-based matching. The matching steps along with the intermediate number of matching journal entries at each stage are illustrated below. The matching steps used criteria of exclusion through which records that yielded in match in one step were excluded from rest of the computations for matching.

Matching step 1: The first matching step involved computing matches based on ISSN and e-ISSN fields. First the records were matched on ISSN and thereafter the remaining ones on e-ISSN. For doing this, the journal lists were partitioned in two sets- those having non-null ISSN value (hereafter referred to as ISSN set) and those with non-null e-ISSN values (hereafter referred to as e-ISSN set). The ISSN set comprised of 13,584 journal entries in Web of Science, 37,780 journal entries in Scopus, and 60,538 journal entries in Dimensions. The e-ISSN set comprised of 12,827 journal entries in Web of Science, 14,203 journal entries in Scopus, and 53,505 journal entries in Dimensions. Both these lists had common journal entries too. To avoid duplicate processing of matching, we removed from e-ISSN set all those records that were already included in the ISSN set. This way, the modified e-ISSN set comprised of 17 records in Web of Science, 1,978 records in Scopus, and 13,428 records in Dimensions.

The subsequent matching on ISSN followed by e-ISSN is done as follows:

a. The entries in the ISSN sets are matched by their ISSN field across all database pairs. This resulted in 12,744 matching records in Web of Science and Scopus, 11,305 matching records in Web of Science and Dimensions, and 23,579 matching records in Scopus and Dimensions.

b. The next step involved matching journal entries in the modified e-ISSN sets of the three databases. Here the entries in the three sets are matched by their e-ISSN values. This resulted in 12 matching records in Web of Science and Scopus, 1,084 matching records in Web of Science and Dimensions, and 8 matching records in Scopus and Dimensions.

c. In the next step, the remaining unmatched records in the ISSN sets after step (a) are matched to modified e-ISSN set with respect to the e-ISSN values. This results in 413 matching records in Web of Science and Scopus, 648 matches in Web of Science and Dimensions, and 43 matching records in Scopus and Dimensions.

d. The remaining ISSN sets are then compared to find any matches on e-ISSN. The ISSN of Web of Science and Scopus have 164 matching e-ISSNs, Web of Science and Dimensions have 763 matching e-ISSNs, and Scopus and Dimensions have 12,246 matching e-ISSNs.

e. Similarly, the modified e-ISSN sets are compared with remaining ISSN set. Web of Science and Scopus have 1 matching e-ISSNs, Web of Science and Dimensions have 3 matching e-ISSNs, and Scopus and Dimensions have 239 matching e-ISSNs.

f. In the last step we did cross matches for the remaining journal entries in both ISSN and e-ISSN sets taken together. The ISSN field in the entries was matched with e-ISSN and vice versa. This was done to address the manual observations that in some records the ISSN and e-ISSN numbers were interchanged in different database lists. This cross matching in the ISSN and e-ISSN sets resulted in 120 matching records in Web of Science and Scopus, 259 matching records in Web of Science and Dimensions, and 999 matching records in Scopus and Dimensions.

Matching Step 2: After the matching of records based on ISSN and e-ISSN fields, we tried to match the remaining records on the journal title field. First an exact title match is done on title fields of records. Thereafter, an inexact match involving cosine similarity is done to process records that have the same journal, spelled or written differently in the three lists. Such cases included journals which are written with '&' in one list and 'and' in the other list, as well as records where one database lists three parts (say part A, B, C) of a journal as a separate entry whereas the other has a single entry for all the three parts taken together. The matching was done as follows:

  a. The remaining records (after first step of matching) in the Web of Science and Scopus databases are matched on the title field of record. First an exact match is done. This resulted in 42 matching records in Web of Science and Scopus. However, 12 records with title match have different publisher information in the two databases. Therefore, they were discarded and we were left with 30 matching records by title field. For Web of Science and Dimensions 180 records matched on title, from which only 144 records have same publisher. In case of Scopus and Dimensions we got 188 title matches out of which 120 records have same publisher.

  b. In the second step of title matching, an inexact match was performed between title fields of the remaining records by computing cosine similarity between them. We considered cosine similarity of 0.9 or higher as an indication of match between two titles. This step resulted in 5 matching records in Web of Science and Scopus, with same publisher name. Therefore, only 5 matching records were considered. Web of Science and Dimensions has 19 records with same publishers out of 22 matches. Similarly, Scopus and Dimensions has 26 records with same publishers out of 56 matches.

The pre-processing and matching steps were executed as above to identify overlapping and unique journal entries across the three databases.

# TABLES

**Table 1: Number of journals and publication records indexed in the three databases**

|  | Web of Science[1] | Scopus | Dimensions |
|---|---|---|---|
| No of journals indexed | 13,610[*] | 40,385[**] | 77,471[***] |
| Approximate number of publication records (article + review) indexed in the three databases (2010-18) | 13,218,007 | 18,058,418 | 28,130,484 |

[1]- Includes SCIE (9,397 journals), SSCI (3,497 journals), AHCI (1,843 journals)

[*] Updated June 2020    [**] Updated June 2020    [***] Updated May 2020

**Table 2: Article Type Distribution in the three databases (2010-18 period)**

| Document Type | Web of Science | Scopus | Dimensions |
|---|---|---|---|
| Article | 12,468,342 | 16,680,987 | 28,130,484 |
| Conference Proceedings Paper (Conferences) | 3,01,619 | 4,393,991 | 2,994,810 |
| Biographical Item | 33,010 | -- | -- |
| Book/ Edited Books | 16 | 163,711 | 116,643 |
| Book Chapter | 40,396 | 1,205,119 | 3,941,124 |
| Book Review | 492,387 | -- | -- |
| Correction | 140,470 | -- | -- |
| Editorial | 854,607 | 577,730 | -- |
| Erratum | -- | 163,707 | -- |
| Letter | 384,090 | 431,041 | -- |
| Meeting Abstract | 2,689,143 | -- | -- |
| Note | -- | 571,858 | -- |
| News Item | 146,838 | -- | -- |
| Preprint | -- | -- | 1,197,813 |
| Retraction | 2,746 | -- | -- |
| Retracted Publication | 3,117 | 3,338 |  |
| Review | 749,665 | 1,377,431 | -- |
| Short Survey | -- | 198,730 | -- |
| Software Review | 340 | -- | -- |
| Conference Review | -- | 52,772 | -- |
| Business Article | -- | 9,073 | -- |
| Data Paper | 1,186 | 3,688 | -- |
| Abstract Report | -- | 1,044 | -- |
| Report | -- | 6 | -- |
| Undefined | -- | 30,524 | -- |
| Monograph | -- | -- | 300,408 |
| WoS others* | 129,615 | -- | -- |

[[*] WoS others include document types- Poetry, Biographical Item, Art Exhibit Review, Record Review, Film Review, Music Performance Review, Fiction Creative Prose, Dance Performance Review, TV Review/ Radio Review, Reprint, Theatre Review, Bibliography, Music Score Review, Database Review, Music Score, Excerpt, Script, Hardware Review, Chronology, Abstract of Published Item, Main Cite and Meeting Summary]

**Table 3: Scheme for mapping subject areas of Scopus and Dimensions to the major subject areas of Web of Science**

| Major subject areas | Databases | | |
|---|---|---|---|
| | **Web of Science** (mapping of 05 major subject areas) | **Scopus** (mapping of 27 subject categories) | **Dimensions** (mapping of 22 divisions of FOR codes) |
| **Arts & Humanities** | Arts & Humanities | Arts & Humanities | 19. Studies in Creative Arts and Writing, 21. History and Archaeology, 22. Philosophy and Religious Studies |
| **Life Sciences & Biomedicine** | Life Sciences & Biomedicine | Medicine, Biochemistry, Genetics and Molecular Biology, Agricultural and Biological Sciences, Environmental Science, Pharmacology, Toxicology and Pharmaceutics, Immunology and Microbiology, Neuroscience, Psychology, Nursing, Health Professions, Veterinary, Dentistry | 11. Medical and Health Sciences, 06. Biological Sciences, 07. Agricultural and Veterinary Sciences, 05. Environmental Sciences |
| **Physical Sciences** | Physical Sciences | Physics and Astronomy, Chemistry, Mathematics, Chemical Engineering, Earth and Planetary Sciences | 03. Chemical Sciences, 02. Physical Sciences, 01. Mathematical Sciences, 04. Earth Sciences |
| **Social Sciences** | Social Sciences | Social Sciences, Business, Management and Accounting, Economics, Econometrics and Finance, Decision Sciences | 17. Psychology and Cognitive Sciences, 16. Studies in Human Society, 15. Commerce, Management, Tourism and Services, 20. Language, Communication and Culture, 13. Education, 14. Economics, 18. Law and Legal Studies |
| **Technology** | Technology | Engineering, Materials Science, Computer Science, Energy | 09. Engineering, 08. Information and Computing Sciences, 10. Technology, 12 Built Environment and Design |

**Table 4: Research output rank, volume and global share of 20 selected countries for 2010-18 period (only document types- 'article' and 'review' included)**

| Country | Web of Science | | | Scopus | | | Dimensions | | |
|---|---|---|---|---|---|---|---|---|---|
| | Rank | Output | Global Share[1] (%) | Rank | Output | Global Share[2] (%) | Rank | Output | Global Share[3] (%) |
| USA | 1 | 3,600,634 | 27.24 | 1 | 4,161,030 | 23.04 | 1 | 4,790,566 | 17.03 |
| China | 2 | 2,282,005 | 17.26 | 2 | 3,213,841 | 17.80 | 2 | 2,637,890 | 9.38 |
| UK | 3 | 908,609 | 6.87 | 3 | 1,239,951 | 6.87 | 3 | 1,450,521 | 5.16 |
| Germany | 4 | 895,656 | 6.78 | 4 | 1,101,725 | 6.10 | 4 | 1,209,059 | 4.30 |
| Japan | 5 | 707,307 | 5.35 | 5 | 859,673 | 4.76 | 5 | 1,164,208 | 4.14 |
| France | 6 | 621,170 | 4.70 | 7 | 772,988 | 4.28 | 6 | 869,547 | 3.09 |
| Canada | 7 | 598,829 | 4.53 | 9 | 681,064 | 3.77 | 8 | 764,264 | 2.72 |
| Italy | 8 | 578,752 | 4.38 | 8 | 691,017 | 3.83 | 9 | 736,789 | 2.62 |
| India | 9 | 529,199 | 4.00 | 6 | 852,896 | 4.72 | 7 | 770,778 | 2.74 |
| Australia | 10 | 524,123 | 3.97 | 11 | 607,092 | 3.36 | 11 | 654,941 | 2.33 |
| South Korea | 11 | 483,207 | 3.66 | 12 | 552,628 | 3.06 | 13 | 558,350 | 1.98 |
| Spain | 12 | 479,277 | 3.63 | 10 | 613,668 | 3.40 | 10 | 658,268 | 2.34 |
| Netherlands | 13 | 350,776 | 2.65 | 15 | 392,326 | 2.17 | 15 | 441,760 | 1.57 |
| Brazil | 14 | 344,012 | 2.60 | 13 | 495,022 | 2.74 | 12 | 616,541 | 2.19 |
| Russia | 15 | 285,171 | 2.16 | 14 | 441,954 | 2.45 | 14 | 443,510 | 1.58 |
| Iran | 16 | 259,748 | 1.97 | 16 | 358,999 | 1.99 | 17 | 308,367 | 1.10 |
| Switzerland | 17 | 257,699 | 1.95 | 18 | 294,936 | 1.63 | 16 | 339,528 | 1.21 |
| Taiwan | 18 | 239,756 | 1.81 | 20 | 262,625 | 1.45 | 21[**] | 260,305 | 0.93 |
| Turkey | 19 | 238,781 | 1.81 | 17 | 299,062 | 1.66 | 20 | 266,517 | 0.95 |
| Sweden | 20 | 235,537 | 1.78 | 21[*] | 261,416 | 1.45 | 19 | 280,988 | 1.00 |

1. Corresponding to total research output of the world in Web of Science (2010-18) = 13,218,007
2. Corresponding to total research output of the world in Scopus (2010-18) = 18,058,418
3. Corresponding to total research output of the world in Dimensions (2010-18) = 28,130,484

[Note: [*]Sweden is not in top 20 in Scopus and [**]Taiwan is not in top 20 in Dimensions. Both these replace Poland, which is 19th in Scopus & 18th in Dimensions, but does not figure in top 20 in Web of Science.]

**Table 5: Country-wise publication record data in the three databases for document types- article + review for the twenty countries during 2010 to 2018**

| Country | Database | Year | | | | | | | | | CAGR (%) |
|---|---|---|---|---|---|---|---|---|---|---|---|
| | | 2010 | 2011 | 2012 | 2013 | 2014 | 2015 | 2016 | 2017 | 2018 | |
| USA | WoS | 350,658 | 366,566 | 379,391 | 393,921 | 401,245 | 409,570 | 422,731 | 433,420 | 443,132 | **2.63** |
| | Scopus | 407,813 | 426,851 | 445,535 | 460,764 | 471,234 | 475,471 | 481,245 | 489,784 | 502,333 | **2.34** |
| | Dimensions | 442,295 | 467,987 | 499,472 | 515,932 | 531,462 | 548,573 | 560,582 | 592,959 | 631,304 | **4.03** |
| China | WoS | 132,277 | 155,639 | 181,145 | 215,468 | 250,866 | 282,355 | 312,391 | 349,245 | 402,619 | **13.17** |
| | Scopus | 230,886 | 253,775 | 286,176 | 332,274 | 366,811 | 388,215 | 416,215 | 444,283 | 495,206 | **8.85** |
| | Dimensions | 154,173 | 209,580 | 235,786 | 274,892 | 299,290 | 303,631 | 324,410 | 385,878 | 450,250 | **12.65** |
| UK | WoS | 83,065 | 87,177 | 91,578 | 97,748 | 98,173 | 104,356 | 110,833 | 115,455 | 120,224 | **4.19** |
| | Scopus | 116,521 | 122,046 | 128,561 | 136,375 | 138,689 | 142,922 | 147,291 | 150,328 | 157,218 | **3.38** |
| | Dimensions | 131,185 | 137,944 | 147,246 | 154,828 | 159,712 | 169,503 | 171,589 | 182,756 | 195,758 | **4.55** |
| Germany | WoS | 83,194 | 88,298 | 93,115 | 97,372 | 99,488 | 102,603 | 107,068 | 111,061 | 113,457 | **3.51** |
| | Scopus | 104,855 | 110,457 | 116,571 | 120,808 | 124,697 | 125,945 | 130,110 | 132,909 | 135,373 | **2.88** |
| | Dimensions | 107,600 | 114,117 | 125,775 | 131,390 | 134,849 | 140,240 | 144,725 | 150,994 | 159,369 | **4.46** |
| Japan | WoS | 74,331 | 76,427 | 77,230 | 78,997 | 77,573 | 77,221 | 79,802 | 82,037 | 83,689 | **1.33** |
| | Scopus | 90,456 | 93,225 | 95,645 | 97,976 | 95,944 | 93,901 | 96,658 | 97,012 | 98,856 | **0.99** |
| | Dimensions | 116,675 | 121,570 | 121,388 | 123,129 | 123,896 | 132,878 | 127,062 | 148,903 | 148,707 | **2.73** |
| France | WoS | 59,163 | 62,254 | 64,755 | 68,087 | 68,784 | 71,307 | 74,499 | 75,925 | 76,396 | **2.88** |
| | Scopus | 75,387 | 78,703 | 82,038 | 85,856 | 87,854 | 87,918 | 91,293 | 91,617 | 92,322 | **2.28** |
| | Dimensions | 78,132 | 84,486 | 93,604 | 98,284 | 100,418 | 101,340 | 103,299 | 103,265 | 106,719 | **3.53** |
| Canada | WoS | 56,568 | 58,965 | 61,916 | 64,955 | 66,305 | 68,491 | 71,061 | 73,717 | 76,851 | **3.46** |
| | Scopus | 64,322 | 67,159 | 71,146 | 74,593 | 77,108 | 78,320 | 79,948 | 82,217 | 86,251 | **3.31** |
| | Dimensions | 67,775 | 71,856 | 78,271 | 81,698 | 84,764 | 88,336 | 90,136 | 96,418 | 105,010 | **4.99** |
| Italy | WoS | 51,927 | 54,792 | 58,310 | 63,067 | 64,687 | 67,094 | 70,673 | 72,735 | 75,467 | **4.24** |
| | Scopus | 60,557 | 64,405 | 69,697 | 75,871 | 78,963 | 80,741 | 84,053 | 86,047 | 90,683 | **4.59** |
| | Dimensions | 62,678 | 66,789 | 73,986 | 79,412 | 82,948 | 87,116 | 90,269 | 93,454 | 100,137 | **5.34** |
| India | WoS | 42,932 | 47,084 | 49,627 | 54,617 | 59,361 | 62,280 | 66,800 | 70,743 | 75,755 | **6.51** |
| | Scopus | 62,988 | 74,866 | 81,488 | 88,295 | 100,039 | 106,483 | 110,813 | 109,514 | 118,410 | **7.27** |
| | Dimensions | 48,458 | 57,155 | 68,581 | 76,653 | 89,100 | 96,346 | 103,942 | 110,649 | 119,894 | **10.59** |
| Australia | WoS | 41,325 | 45,594 | 49,541 | 55,129 | 58,817 | 63,269 | 66,999 | 69,793 | 73,656 | **6.63** |
| | Scopus | 49,794 | 53,657 | 58,148 | 65,215 | 69,862 | 73,276 | 75,981 | 78,144 | 83,015 | **5.84** |
| | Dimensions | 49,804 | 55,478 | 61,450 | 67,313 | 72,652 | 78,956 | 81,473 | 89,896 | 97,919 | **7.80** |
| South Korea | WoS | 40,591 | 45,057 | 49,284 | 51,764 | 54,814 | 57,995 | 59,718 | 60,624 | 63,360 | **5.07** |
| | Scopus | 44,911 | 48,905 | 54,576 | 59,502 | 64,645 | 67,789 | 69,385 | 70,414 | 72,501 | **5.47** |
| | Dimensions | 44,356 | 50,029 | 56,753 | 60,189 | 62,923 | 67,931 | 69,677 | 71,442 | 75,050 | **6.02** |
| Spain | WoS | 42,297 | 46,532 | 50,193 | 52,912 | 54,048 | 55,207 | 57,487 | 59,032 | 61,569 | **4.26** |
| | Scopus | 54,127 | 59,494 | 64,054 | 67,736 | 70,260 | 70,501 | 73,499 | 75,483 | 78,514 | **4.22** |
| | Dimensions | 55,223 | 59,828 | 66,737 | 70,689 | 74,613 | 78,116 | 79,870 | 83,760 | 89,432 | **5.50** |
| Netherlands | WoS | 32,075 | 33,801 | 36,481 | 38,444 | 39,029 | 40,317 | 42,341 | 43,305 | 44,983 | **3.83** |
| | Scopus | 36,182 | 37,991 | 41,307 | 43,525 | 44,839 | 45,171 | 46,631 | 47,332 | 49,348 | **3.51** |

|  |  |  |  |  |  |  |  |  |  |  |
|---|---|---|---|---|---|---|---|---|---|---|
|  | Dimensions | 38,140 | 41,476 | 45,537 | 48,750 | 49,784 | 51,232 | 52,691 | 54,411 | 59,739 | **5.11** |
| Brazil | WoS | 26,136 | 29,301 | 32,915 | 35,186 | 37,428 | 40,044 | 44,436 | 47,497 | 51,069 | **7.73** |
|  | Scopus | 40,860 | 44,671 | 49,159 | 51,910 | 55,401 | 57,224 | 61,545 | 64,964 | 69,288 | **6.04** |
|  | Dimensions | 47,548 | 51,171 | 57,473 | 61,044 | 66,035 | 73,587 | 79,269 | 86,546 | 93,868 | **7.85** |
| Russia | WoS | 26,035 | 27,509 | 26,775 | 28,414 | 29,867 | 34,061 | 35,611 | 37,358 | 39,541 | **4.75** |
|  | Scopus | 32,813 | 35,834 | 35,324 | 40,368 | 45,701 | 53,795 | 61,267 | 65,085 | 71,767 | **9.08** |
|  | Dimensions | 33,170 | 35,203 | 35,391 | 38,497 | 43,472 | 52,960 | 57,325 | 68,385 | 79,107 | **10.14** |
| Iran | WoS | 17,265 | 22,073 | 24,767 | 26,469 | 27,941 | 29,917 | 33,843 | 37,457 | 40,016 | **9.79** |
|  | Scopus | 22,667 | 31,600 | 34,656 | 37,238 | 40,495 | 41,078 | 47,531 | 50,148 | 53,586 | **10.03** |
|  | Dimensions | 16,751 | 23,581 | 27,355 | 30,602 | 34,063 | 37,067 | 41,043 | 46,455 | 51,450 | **13.28** |
| Switzerland | WoS | 22,183 | 23,980 | 25,845 | 27,445 | 28,453 | 29,755 | 31,925 | 33,703 | 34,410 | **5.00** |
|  | Scopus | 25,566 | 27,809 | 30,272 | 32,033 | 33,569 | 34,277 | 35,822 | 37,380 | 38,208 | **4.57** |
|  | Dimensions | 27,708 | 30,708 | 34,138 | 36,424 | 37,639 | 39,936 | 40,949 | 44,695 | 47,331 | **6.13** |
| Taiwan | WoS | 24,900 | 27,261 | 27,688 | 28,197 | 27,680 | 26,617 | 26,433 | 25,460 | 25,520 | **0.27** |
|  | Scopus | 27,499 | 29,539 | 30,857 | 31,540 | 30,730 | 29,102 | 28,620 | 27,379 | 27,359 | **-0.06** |
|  | Dimensions | 25,029 | 28,845 | 29,553 | 31,089 | 29,971 | 29,927 | 28,206 | 28,385 | 29,300 | **1.77** |
| Turkey | WoS | 21,608 | 22,575 | 24,153 | 25,624 | 26,484 | 28,409 | 30,807 | 29,145 | 29,976 | **3.70** |
|  | Scopus | 27,190 | 28,972 | 30,475 | 32,949 | 33,860 | 35,698 | 38,153 | 35,269 | 36,496 | **3.32** |
|  | Dimensions | 20,635 | 21,927 | 25,101 | 29,072 | 30,970 | 34,706 | 35,140 | 33,495 | 35,471 | **6.20** |
| Sweden | WoS | 20,451 | 21,612 | 23,397 | 25,082 | 26,156 | 27,441 | 29,386 | 30,519 | 31,493 | **4.91** |
|  | Scopus | 22,865 | 24,186 | 26,083 | 28,001 | 29,793 | 30,974 | 32,138 | 33,091 | 34,285 | **4.60** |
|  | Dimensions | 23,138 | 24,824 | 27,789 | 29,454 | 31,354 | 33,326 | 34,873 | 36,897 | 39,333 | **6.07** |

# FIGURES

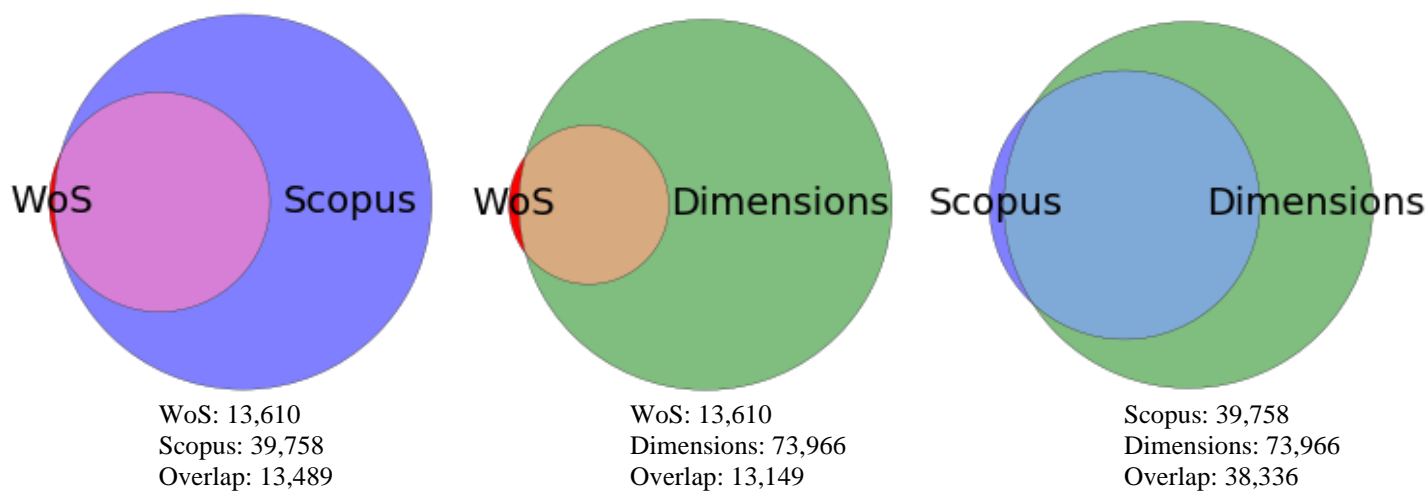

WoS: 13,610
Scopus: 39,758
Overlap: 13,489

WoS: 13,610
Dimensions: 73,966
Overlap: 13,149

Scopus: 39,758
Dimensions: 73,966
Overlap: 38,336

Figure 1: Pair-wise journal coverage overlap of the three databases

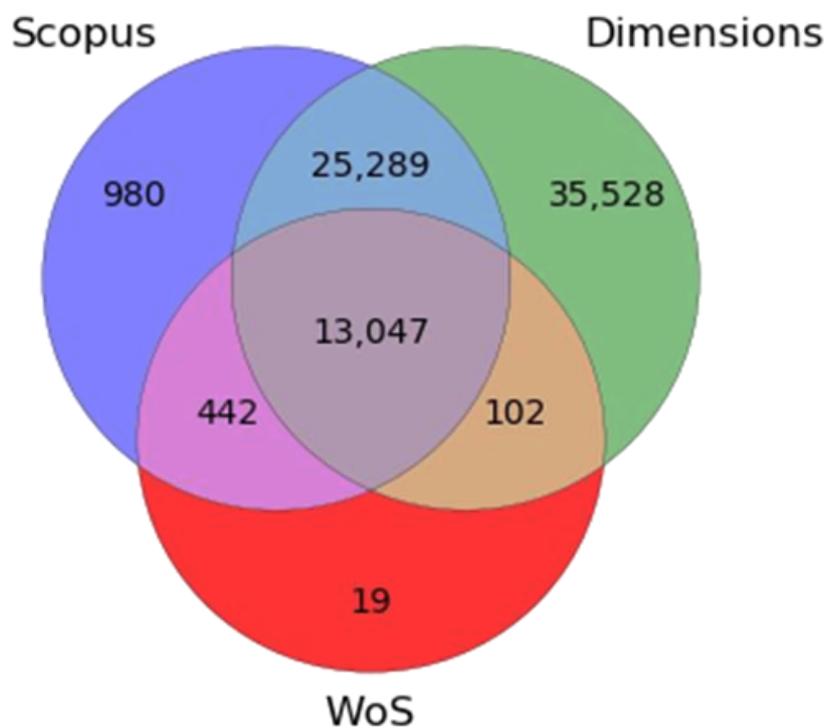

Figure 2: Journal coverage overlap of all the three databases

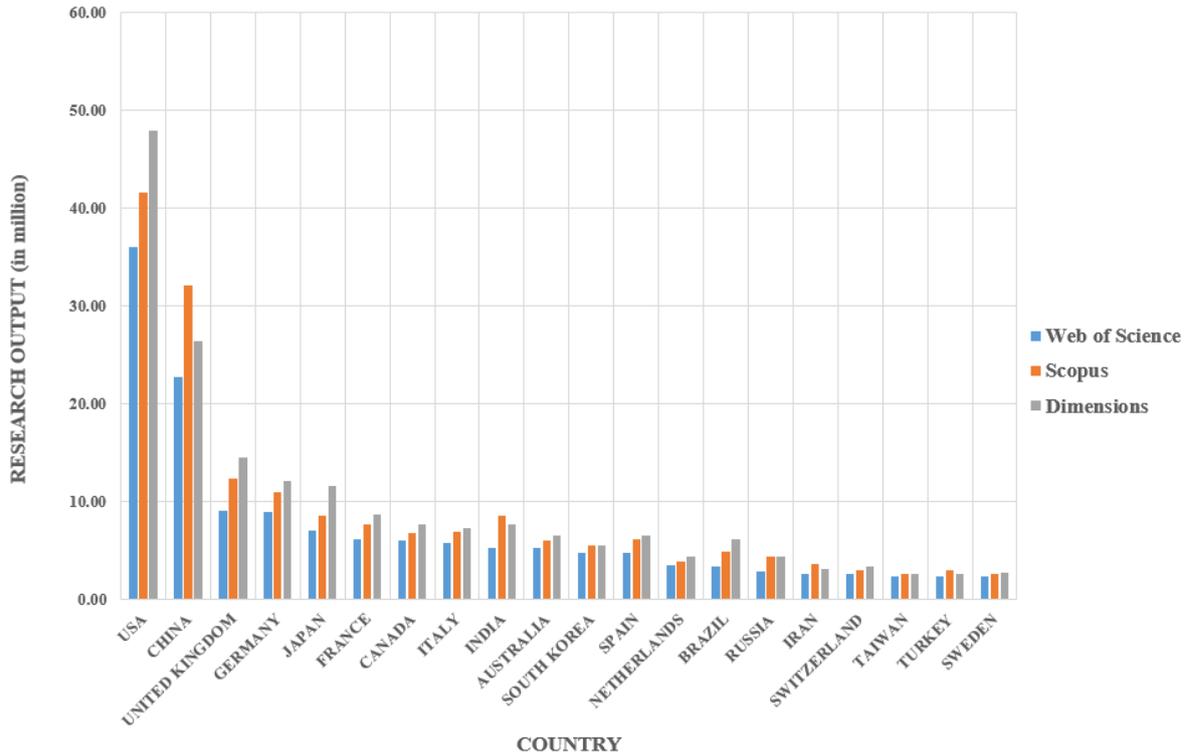

Figure 3: Research output (article + review) of the 20 countries in the three databases (2010-18)

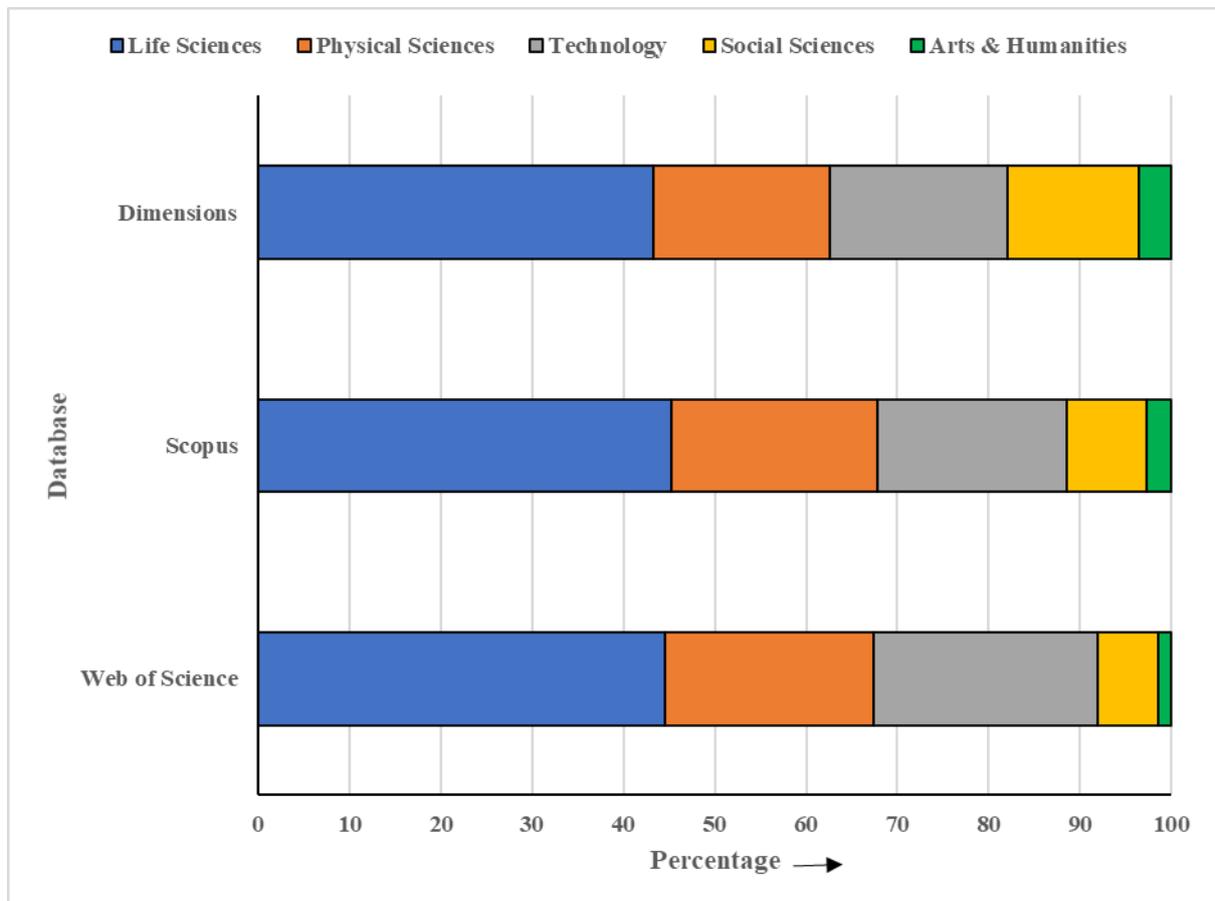

Figure 4: Distribution of combined research output (article + review) of the 20 countries (2010-18) in the three databases, according to WoS major subject area classification

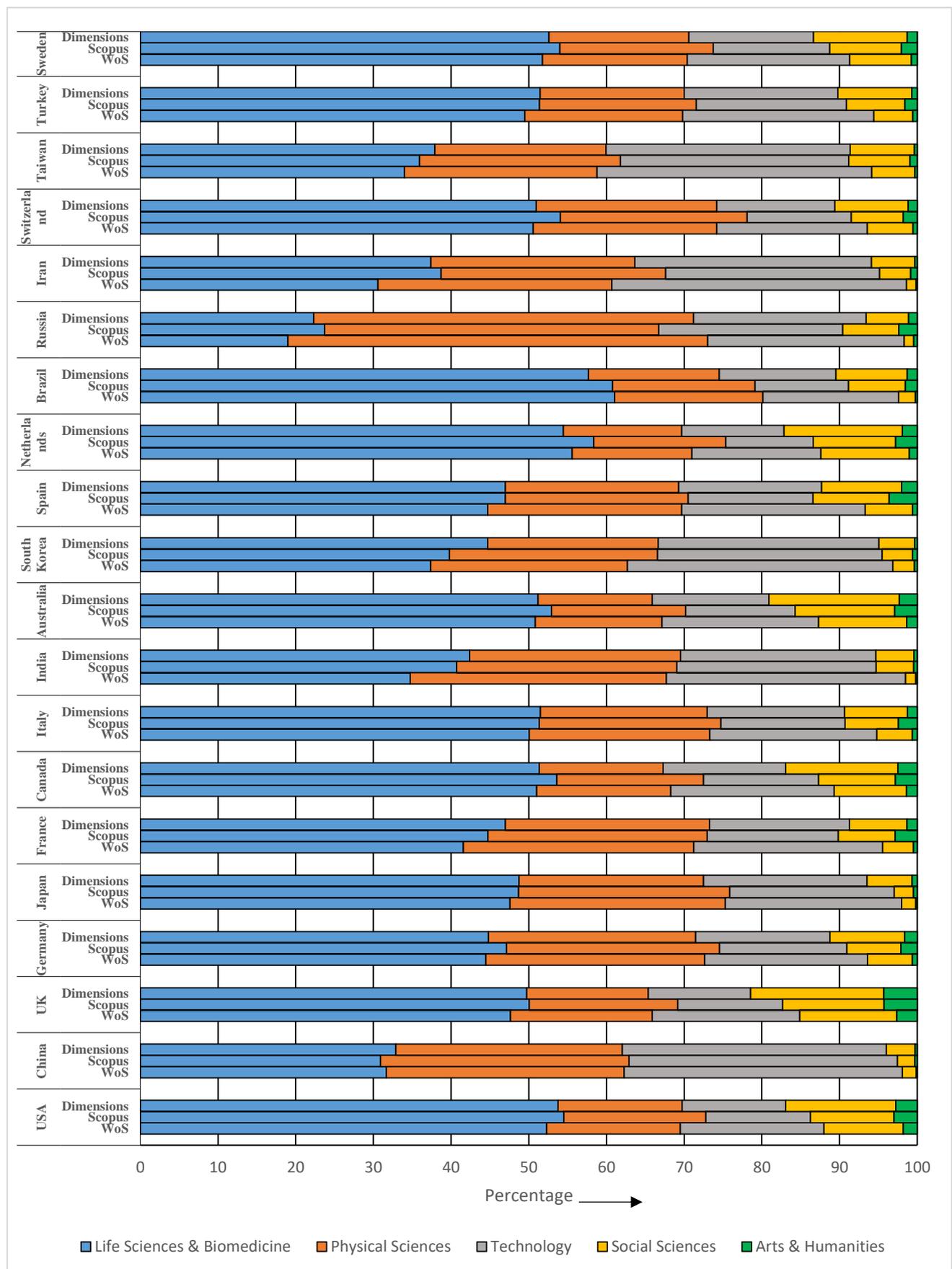

Figure 5: Distribution of individual research outputs (article + review) of the 20 countries (2010-18) in the three databases, according to WoS major subject area classification